\documentclass[12pt]{article}
\usepackage{subfig}
\usepackage{epsfig}
\usepackage{graphicx}
\usepackage{multicol}
\usepackage{amsfonts}
\usepackage{amsmath,amssymb}
\usepackage{enumerate}
\numberwithin{equation}{section}
\usepackage{cite}

\newcommand{\ben}{\begin{enumerate}}
\newcommand{\een}{\end{enumerate}}
\newcommand{\spa}{\phantom{s}}
\newcommand{\vp}{\varphi}
\newcommand{\bea}{\begin{eqnarray}}
\newcommand{\eea}{\end{eqnarray}}
\newcommand{\be}{\begin{equation}}
\def\bel#1{\begin{equation} \label{#1}}
\newcommand{\ee}{\end{equation}}
\newcommand{\bi}{\begin{itemize}}
\newcommand{\ei}{\end{itemize}}
\newcommand{\ba}{\begin{align}}
\newcommand{\ea}{\end{align}}
\newcommand{\comments}[1]{}

\def\pref#1{(\ref{#1})}
\def\hf{\frac12}

\def\mvp{m_{\varphi}}
\begin{document}

\begin{titlepage}
\vskip 1 cm
\begin{center}
{\Large \bf   Inflationary Predictions and Moduli Masses
}
\vskip 1.5cm  
{ 
\sc{Kumar Das$^{*}$, Koushik Dutta$^{*}$, Anshuman Maharana$^{\dagger}$}
\let\thefootnote\relax\footnotetext{ \hspace{-0.5cm} E-mail: {$\mathtt{kumar.das@saha.ac.in, koushik.dutta@saha.ac.in, anshumanmaharana@hri.res.in} $}}
}
\vskip 0.9 cm

{\textsl{$^{*}$Theory Division, \\
 Saha Institute of Nuclear Physics, \\
 1/AF Salt Lake,  \\
 Kolkata - 700064, India.}
}
\vskip 0.6 cm
{\textsl{
$^{\dagger}$Harish Chandra Research Intitute, \\
Chattnag Road, Jhunsi,\\
Allahabad -  211019, India.\\}
}
%
%
\end{center}

\vskip 0.6cm

\begin{abstract}
\vskip 0.5 cm
{
  A generic feature of inflationary models in supergravity/string constructions
is vacuum misalignment for the moduli fields. The  associated production of moduli
particles leads to an epoch in the post-inflationary history in which the  energy
density is dominated by cold moduli particles. This modification of the post-inflationary
history implies that the preferred range for the number of e-foldings between horizon exit
of the modes relevant for CMB observations and the end of inflation $(N_k)$ depends on 
moduli masses. This in turn implies that the precision CMB observables $n_s$ and $r$ are
sensitive to moduli masses. We analyse this sensitivity for some representative models of
inflation and find  the effect to be highly relevant for confronting inflationary models
with observations.}
\noindent

\end{abstract}

\vspace{3.0cm}

\end{titlepage}
\pagestyle{plain}
\setcounter{page}{1}
\newcounter{bean}
\baselineskip18pt
%


%
%

\section{Introduction}

   Precision measurements of the cosmic microwave background (CMB) have put the inflationary paradigm as the leading candidate for a theory of early universe cosmology. The data is in perfect agreement with the basic qualitative predictions of inflation i.e. an approximately scale invariant and adiabatic power spectrum. Upcoming observations are expected to probe the CMB with an even greater accuracy and provide us
information regarding the strengths of the tensor to scalar ratio and non-gaussianities.

     On the theoretical front, there are many challenges. The inflationary slow
roll conditions are ultraviolet sensitive; 
we should embed models of inflation in a quantum theory of gravity. In this light, an
important direction of research is study of the effects that can arise
as a result of ultraviolet completion of inflationary models. String theory provides
a setting where one can hope to carry out a systematic study of such effects.

A generic feature of supergravity/string models are the  moduli fields. The vacuum expectation
value of moduli fields set the strength and form of the low energy effective action of
string models, hence moduli fields play a central role in string phenomenology.
In the context of inflationary model building, a knowledge of the moduli potential
and the couplings between the inflaton and the moduli are necessary to address the $\eta$
problem. There has been an extremely useful interplay between studies of moduli stabilisation and inflationary model building in string theory, see for e.g. \cite{R1,R2,R3}. In this paper, we will
examine the sensitivity of precision CMB observables -- the spectral tilt $(n_s)$ and 
the  tensor to scalar ratio $(r)$ to the mass of the lightest modulus field.

     Given a model of inflation, one can express $n_s$ and $r$ in terms of the 
number of e-foldings between horizon exit of the modes relevant for CMB observations
and  the end of inflation $(N_k)$. Predictions for $n_s$ and $r$ are then made by
using the ``preferred range" of $N_k$  (usually taken to be 50 to 60) in these formulae. 
The preferred range for $N_k$ is determined as follows -- the inflationary paradigm
relates the strength of temperature fluctuations in the CMB to the energy density of the universe at the time of  horizon exit $(\rho_k)$ and the tensor to scalar ratio $(r)$. Thus, in the
context of inflationary models the measurement of the temperature fluctuations of the CMB determines the energy density of the universe at the time of horizon exit (modulo $r$). 
The CMB measurements via determination of the Hubble constant also provide us with
the energy density of the universe today $(\rho_0)$. This implies a consistency condition for any theoretical proposal for the  history of the universe between the time of horizon exit of the modes relevant for CMB observations and today - the ``early time" energy density $\rho_k$ should precisely evolve to the energy density today $\rho_0$. In the standard model of
cosmology, the history of the universe between horizon exit and today consists of the following epochs - inflation, reheating, radiation domination and  matter domination. Applying the above mentioned consistency condition to this cosmological timeline, 
along with  the assumption that the reheating epoch is generic gives
$N_k$ to be in the range of 50 to 60 (see for e.g. \cite{planck14}).

          From the very early days of inflationary model building in supergravity
 it was realised that a generic implication of having moduli fields is a non-standard post-inflationary cosmological timeline \cite{cmp,ccmp, cmmp, fq, douglas, bobby} (often referred to as the modular cosmology timeline).  The modular cosmology timeline sets
in whenever there is a modulus of post-inflationary mass\footnote{Curvature couplings imply that the mass
of  a modulus field can be significantly different during the inflationary and post-inflationary epochs.} $(\mvp)$ below the Hubble scale 
during inflation. Such a  modulus gets displaced from its post-inflationary minimum due to vacuum misalignment during the inflationary epoch. This misalignment
is responsible for the non-standard post-inflationary cosmological timeline (which we will
describe in detail in section 2). The distinguishing feature of this timeline is an epoch  in which the energy density of the universe is dominated by cold moduli particles produced  as a result of the misalignment. The duration of this epoch is governed by the lifetime of the modulus (which is typically set by the mass of the modulus as moduli decay via their Planck suppressed interactions). The universe reheats for a second time with the decay of the modulus (the first reheating being associated with the decay of the inflaton), after which the history is thermal\footnote{The successes of big bang nucleosynthesis imply that the decay of the modulus
has to take place before nucleosynthesis.} . Reference \cite{dm} derived the  consistency
condition described in the previous paragraph for the modular cosmology timeline.
With the assumption of generality on the reheating epoch,
the consistency condition turns out to be a relationship between $N_k$ and the number
of e-foldings of the universe during the epoch in which the cold moduli
particles dominate the energy density $(N_{\rm mod})$
\bel{rela}
  50   \lesssim N_k + {1 \over 4} N_{\rm mod}  \lesssim 60.
\ee

      In this paper our goal is to explore in detail the phenomenological implications
of the above relation. The relation can be thought of as implying a shift in the 
central value of the preferred number of e-foldings\footnote{See \cite{Liddle} for a systematic
discussion of various effects that can affect the preferred range for $N_k$.}
\bel{shift}
  55 \to 55 - {1 \over 4} N_{\rm mod}.
\ee
$N_{\rm mod}$ can be expressed in terms of  the mass of the lightest modulus 
$(m_{\varphi})$ in the compactification and the ``initial displacement" of the modulus in Planck units $(Y)$, that occurs as a result of misalignment \cite{dm} 
\bel{mod}
   N_{\rm mod} \approx 
   {4 \over 3} \ln  \bigg( { \sqrt{16 \pi} M_{\rm pl} Y^{2}  \over m_{\varphi} } \bigg).
\ee
 Typically, in supergravity models the initial displacement $Y$ is of  $\mathcal{O}(1)$
\cite{mismatch, mismatchb, DineR, DineRT, quan}. Thus the shift in the central value of $N_k$ is essentially determined by the 
mass of the modulus. 

   Recall that given a model of inflation, the predictions for $n_s$ and $r$ are
made by expressing $n_s$ and $r$ in terms of $N_k$; and then making use of the preferred
range for $N_k$ in these expressions. As discussed above, in supergravity models
the central value of the preferred range for $N_k$ depends on the mass of the lightest modulus. Hence, {{\it the predictions for}} $n_s$ {\it and} $r$ {\it{of an inflationary model are  sensitive to its embedding into a supergravity/string compactification.}} In this paper, we will study  this sensitivity for   some  representative models of inflation ($m^2 \chi^{2}$ \cite{Linde}, axion monodromy \cite{mono}, natural inflation \cite{KF} and the Starobinsky model \cite{Star}). Motivated by the varied spectra of phenomenologically viable supergravity models
 we will treat the mass of the lightest modulus $(m_{\varphi})$ as a parameter.  At any given value of  $m_\varphi$, the preferred range of $N_k$
can be computed using \pref{shift} and \pref{mod}, the predictions for $n_s$ and $r$
can then be made.

   We analyse our results in the context of   {\sc{planck}} 2015 data \cite{planck15}. The implications are very interesting -- the changes in inflationary predictions can significantly affect the scorecard
for  models. For example, a modulus of mass below $10^{5} \phantom{a} {\rm{TeV}}$ can bring the
axion monodromy model within the $1\tiny{-}\sigma$ region for $n_s$ and $r$ based on observations of TT modes at low P. For many models, the effect is important even for very heavy moduli
$(m_{\vp} \approx 10^8 \spa {\rm TeV})$.
  
    This paper is organised as follows. In section 2, we review modular cosmology
and some of the results of \cite{dm} (in particular, the relation between $N_k$ and $N_{\rm mod}$, i.e. equation
\pref{rela} of the introduction). In section 3, we discuss the predictions for $n_s$ and $r$
of some representative models of inflation as a function of $m_{\varphi}$ (the mass of the lightest
modulus). In section 4, we analyse in the context of {\sc{planck}} 2015 data a bound on moduli masses obtained  using \pref{rela} in \cite{dm}. We conclude in section 5 with some discussion.
We review the generation  of density perturbations in modular cosmology in the Appendix.

\section{Review}

\subsection{Modular Cosmology}
\label{Modcos}

   At tree level,  string compactifications  have massless scalar fields  
which interact via Planck suppressed interactions (the moduli). Moduli
acquire masses from sub-leading effects, their masses are typically well
below the string scale and hence moduli are part of the low energy effective action.

    Moduli fields usually have curvature couplings; this makes their masses and potential dependent
on the expectation value of the inflaton. As a result, the minimum of the potential for
a modulus of post-inflationary mass less than Hubble  during inflation $(m_{\vp} < H_{\rm infl})$ is different during the inflationary and post-inflationary epochs -- such a modulus finds itself displaced from its post-inflationary minimum at the end of inflation. This ``initial displacement" is typically of the order of $M_{\rm pl}$ \cite{mismatch, mismatchb, DineR, DineRT, quan}.

    As discussed in the introduction, this ``misalignment" implies a non-standard cosmological timeline.
We briefly review this timeline and refer the reader to \cite{cmp,ccmp, cmmp, fq, douglas, bobby} for a more complete
discussion. Let us begin by describing the case when there is a single modulus whose post-inflationary mass
$m_{\vp}$ is below the Hubble scale during inflation.
At the end of inflation the universe reheats, the energy density  associated with the inflaton
gets converted to radiation. At this stage, the energy density of the universe consists of two components -
radiation, and the energy associated with the modulus displaced from its minimum\footnote{Since $m_{\vp} <
H_{\rm infl}$, right after reheating the energy density associated with radiation dominates over the energy density associated with the displaced modulus.}. Also,
the high value of the Hubble friction keeps the modulus pinned at its initial displacement. As the
universe cools, the Hubble constant drops. When the Hubble friction falls below the mass of the modulus, the modulus begins to oscillate about its post-inflationary minimum. With this, the associated energy density dilutes as matter i.e. much slower than
that of the radiation. Eventually the energy density associated with the modulus dominates the energy
density of the universe; the universe enters into the epoch of modulus domination. This epoch lasts until
the decay of the moduli particles. The modulus decays via its Planck suppressed interactions, the
characteristic lifetime is set by the mass of the modulus \cite{fq, douglas, bobby}
\bel{life}
  \tau_{\rm mod}  \approx {16 \pi M_{\rm pl}^2 \over m_{\vp}^3 }.
\ee
The universe reheats for a second time after the decay of the modulus, after which the history is thermal.
In summary, the modular cosmology timeline consists of the following epochs - inflation, reheating (associated with  inflaton decay), radiation domination, modulus domination, reheating (associated with modulus decay),
radiation domination, matter domination and finally the present epoch of acceleration.

   In models with multiple moduli with post inflationary mass below Hubble during inflation, there are multiple epochs of modulus domination and reheating associated with the moduli. In cases where there is a 
separation of scale between the  mass of the lightest modulus  and the mass of other moduli the lightest modulus outlives the others and sets the time scale for the epoch of modulus domination. The dynamics of the
system can be effectively described by a model with a single modulus; with the effect of the heavy moduli
being incorporated in the reheating epoch after inflation (the $m_{\vp}^{-3}$ dependence of the lifetime
\pref{life} implies that this effective description can be useful even for a moderate separation between the
mass of the lightest moduli and the heavier ones). In models in which there is no distinct lightest modulus the dynamics is more complicated to analyse; this was discussed briefly in \cite{dm}. We will confine
ourselves to situations in which there is a distinct lightest modulus in this paper.

     An important constraint on the modular cosmology timeline arises from the successes big bang of nucleosynthesis. The reheat temperature after the decay of the modulus is given by the formula
\begin{equation}
   T_{\rm rh2} \approx m_{\vp}^{3/2} M^{-1/2}_{\rm pl}.
\end{equation}
Demanding that this is sufficiently high for successful nucleosynthesis gives the cosmological
moduli problem (CMP) bound \cite{cmp,ccmp, cmmp, Kawasaki} on moduli masses
\bel{bound}
    m_{\vp} > 30 \spa {\rm TeV}.
\ee
As mentioned in the introduction, reference \cite{dm} obtained a bound for moduli masses based
on the requirement that density perturbations of the right magnitude are generated in modular cosmology.
The bound is a function of $N_k$, and can be significantly stronger than the CMP bound \pref{bound}.
In section 4, we will explore this bound in detail in the context of  {\sc{planck}}  2015 results
\cite{planck15}.

\subsection{Relating the Number of e-foldings between Horizon Exit and end of Inflation to Mass of the Modulus}
\label{efoldmass}

    In this section we briefly review the results of \cite{dm} relevant for our analysis. Our focus will be on models in which adiabatic perturbations
are generated as a result of quantum fluctuations during the inflationary epoch. The strength of the
inhomogeneities generated is given by  
$$
   A_s = {2 \over 3 \pi^2 r} \bigg( {\rho_k \over M_{\rm pl}^{4} } \bigg),
$$
where $A_s$ is the amplitude of the scalar perturbations, $\rho_k$ the energy density of the universe
at the time of horizon exit  and $r$ the tensor to scalar ratio. We review the details of generation
of density perturbations in the context of modular cosmology in Appendix A. The scalar amplitude $A_s$ is constant to a very good approximation until the point of horizon re-entry  and can be related to the strength of temperature fluctuations in the CMB.
Thus the measurement of the strength
of temperature fluctuations gives us the value of the energy density of the universe at the time of
horizon exit (modulo r). CMB observations also give us the value of the  energy density today $(\rho_0)$ via
determination of the Hubble constant. Thus any theoretical proposal for the history of the universe between horizon exit and the present epoch must be such that $\rho_k$ evolves to  $\rho_0$. Reference \cite{dm}
applied this consistency condition to the modular cosmology timeline described in section \ref{Modcos}.
This gave the relation\footnote{ The analogous relation for the standard cosmological timeline is
 \bea
\label{general3}
 N_k   + {1 \over 4}( 1 - 3w_{\rm re}) N_{\rm re}  \approx 55.43  +\frac{1}{4}\ln r    + { 1 \over 4 } \ln \left(\frac{{ \rho_{ k}}} {\rho_{\rm end}} \right).
\eea
}
 \bea
\label{general3}
 N_k +{1 \over 4}{N_{\rm mod} }  + {1 \over 4}( 1 - 3w_{\rm re1}) N_{\rm re1}  +  {1 \over 4}( 1 - 3w_{\rm re2}) N_{\rm re2} \approx 55.43  +\frac{1}{4}\ln r    + { 1 \over 4 } \ln \left(\frac{{ \rho_{ k}}} {\rho_{\rm end}} \right),
\eea
where $N_k$ is the number of e-foldings between horizon exit of the modes relevant for CMB observations
and the end of inflation, $N_{\rm mod}$ is the number of e-foldings that the universe undergoes during the
epoch of modulus domination, $w_{\rm re1}$ is the effective equation of state during the first reheating epoch
(associated with inflaton decay), $N_{\rm re1}$ is the number of e-foldings during the first reheating epoch,
$w_{\rm re2}$ is the effective equation of state during the second reheating epoch
(associated with modulus decay), $N_{\rm re2}$ is the number of e-foldings during the second reheating epoch,
$r$ is the tensor to scalar ratio, $\rho_k$ the energy density at the time of horizon exit and
$\rho_{\rm end}$ the energy density at the end of inflation. The number of e-foldings of modulus 
domination was found to be
\bel{Modd}
   N_{\rm mod} \approx 
   {4 \over 3} \ln  \bigg( { \sqrt{16 \pi} M_{\rm pl} Y^{2}  \over m_{\varphi} } \bigg)
\ee
where as in \pref{mod}, Y is the initial displacement of the modulus from its post-inflationary
minimum in Planck units.  Equation \pref{general3} can be used to obtain the ``preferred range"
of $N_k$ for modular cosmology. A discussion of the analogous analysis for the standard cosmological
timeline can be found in \cite{planck14}. Making the same generality assumptions regarding the reheating
epoch, change in the energy density of the universe during inflation and the scale of inflation 
as in section 2.3 of \cite{planck14}, equation \pref{general3} gives the preferred range 
for $N_k$ to be
\bel{pr}
    \bigg(55 - {1 \over 4}{N_{\rm mod}} \bigg) \pm 5.
\ee
Note that this can be thought of as lowering of the central value of the preferred range of
$N_k$ by $N_{\rm mod}/4$. As mentioned earlier, there are general arguments  which imply $Y$ 
is an $\mathcal{O}(1)$ quantity\footnote{These expectations have been borne out in explicit constructions
of inflationary models in string compactifications, see for e.g. \cite{finer}.}. Thus the shift
in the central value of $N_k$ is essentially determined by $m_{\vp}$.

   Before ending this section we would like to emphasise that the relation \pref{general3} and expression
\pref{pr} are valid only if the post inflationary mass of the modulus $m_{\vp}$ is below Hubble during inflation. If the  post-inflationary mass of the lightest modulus is well above Hubble during
inflation then the misalignment mechanism is not operational and the preferred range is $55 \pm 5$.

\section{Implications for Inflationary Models}
    In this section, we will study the phenomenological implications of the results described in
section \ref{efoldmass} for some representative models of inflation. Given the diverse spectra of
 phenomenologically viable supergravity models we will treat $m_{\varphi}$ as a phenomenological
parameter in our analysis. The central value of $N_k$ \pref{pr} also depends on $Y$, the initial
displacement in Planck units. Given a model of inflation (embedded in a supergravity construction)
this can be computed explicitly; see \cite{DineR, DineRT} for an outline of the method. On the other hand,
as described earlier there are various arguments based on general principles which imply Y is of
$\mathcal{O}(1)$. Guided by these arguments, in what follows we will work with $Y = 1/10$. 
Note that the shift in the central value of $N_k$ decreases with Y. So our choice of $Y = 1/10$ can be
considered conservative; but this ensures better control over the effective field theory.

We will focus on four benchmark models of inflation - $ V(\chi) = \hf m^2 \chi^2$ \cite{Linde} (we will denote the inflaton by $\chi$),
axion monodromy i.e $ V(\chi) =\hat{m}^{10/3} \chi^{2/3}$ \cite{mono},  natural (pNGB) inflation \cite{KF} and 
the Starobinsky model \cite{Star}. Let us record $n_s$ and $r$ as a function of $N_k$ for each of these models 
\begin{itemize}
\item $m^{2} \chi^2$:   $n_s = 1 - 2/N_k,$ \spa \spa $r = 8/N_k$ 
\item Axion monodromy: $n_s = 1-  4/(3N_k) ,$ \spa \spa $r = 8/(3N_k)$
\item Natural inflation: $n_s= 1 - \bigg[\frac{M_{\rm pl}}{f}\bigg]^2 \bigg[\frac{1+ \frac{e^{-x}}{p}}{1-\frac{e^{-x}}{p}}\bigg]$, \spa \spa $ r= 8\bigg[\frac{M_{\rm pl}}{f}\bigg]^2 \bigg[\frac{\frac{e^{-x}}{p}}{1-\frac{e^{-x}}{p}}\bigg]$ with $p=1+\frac{M_{\rm pl}^2}{2 f^2},$ \phantom{Natural Inflation:::}$x=\frac{N_k M_{\rm pl}^2}{f^2}$ where $f$ is the axion decay constant.%
\item Starobinsky model: $n_s = 1 - 2/N_k,$ \spa \spa $r = 12/N_k^2$

\end{itemize}

\begin{figure}[h!]
\centering
\includegraphics[width=13cm]{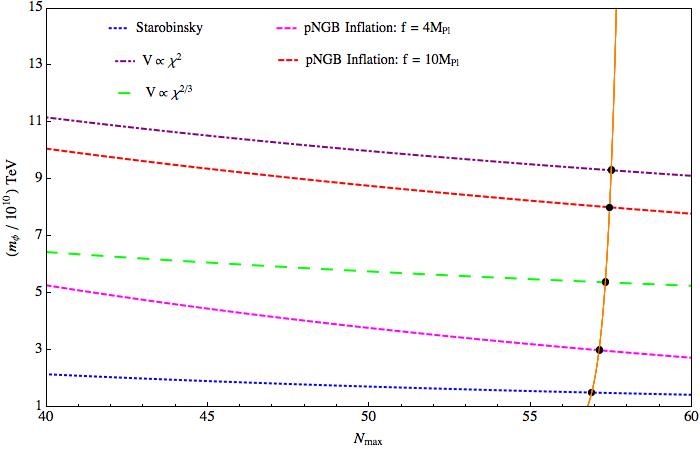}
\caption{Numerical solution for the condition $H_{\rm infl} > \mvp $. The solid curve is a  plot of $\mvp$ as a function of $N_{\rm max}$ as given by \pref{Max}. The dashed curves are plots of the left hand side of 
\pref{condition} as a function of $N_{\rm max}$ for various models. }
\label{upper}
\end{figure}

    The change in the preferred range of $N_k$  \pref{pr} occurs if  $m_{\vp}$ is less than 
Hubble during inflation. We begin by implementing this condition for each of the models. The Hubble constant at the time
of horizon exit is
\bel{Hubexit}
   H_k = { \pi \over \sqrt{2} } (A_s r)^{1/2} M_{\rm pl}
\ee
Note that the
right hand side of \pref{Hubexit} depends on $m_{\varphi}$; since $r$  is determined by $N_k$
and the preferred range for $N_k$ depends on $\mvp$. Also,  $r$ decreases with an increase in $N_k$. Therefore, the condition can be implemented over the entire preferred range by requiring that it holds  for the maximum 
value of $N_k$ 
\bel{Max}
  N_{\rm max} =  60 - {1 \over 3} \ln \bigg( { \sqrt {16 \pi} M_{\rm pl} Y^2 \over m_{\varphi} } \bigg).
\ee
Thus we want to impose the condition
\bel{condition}
   { \pi \over \sqrt{2} } (A_s r[N_{\rm max}])^{1/2} M_{\rm pl} > \mvp
\ee
with $N_{\max}$ as given by \pref{Max}. We solve for this condition numerically in the plot shown in Figure \ref{upper}.
\vspace{1cm}
The condition is most stringent for the Starobinsky model, for which the right hand side and left hand
side of \pref{condition} are equal for  $m_{\varphi} \approx 1.5 \times 10^{10} \spa \rm{TeV}$.  We will
be conservative and study the implications of the shift in the central value of $N_k$ if the
mass of the modulus is at least two orders of magnitude below this i.e $m_{\vp} < 10^{8} \spa \rm{TeV}$ (this  value will be used for all models).

 As described in section \ref{Modcos},  successful nucleosynthesis requires $m_{\vp} > 30 \spa {\rm TeV}$.
We will use this consideration to set the lower value of $m_{\vp}$  in our analysis. In summary,  we will use
the range
\bel{range}
   10^2\spa {\rm TeV} < \mvp < 10^{8} \spa {\rm TeV}
\ee
to study the effects of the epoch of modulus domination on inflationary predictions.

    We now have all the ingredients necessary to compute the predictions for $n_s$ and $r$. For 
$\mvp$  in the range given by \pref{range} the preferred range for $N_k$ is given by \pref{pr}. On the other
hand, if the mass of the modulus is greater than Hubble during inflation, the preferred range is
50-60.  We  compute the predictions for $n_s$ and $r$ for $\mvp = 10^{3}, 10^{6} \spa  \rm{and} \spa  10^{8} \spa \rm{TeV}$.
\begin{figure}[!h]
    \centering
    \subfloat{{\includegraphics[width=7.4cm]{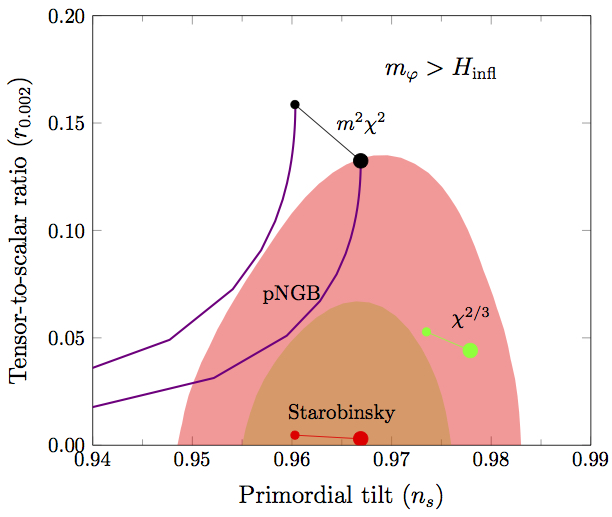} }}%
    \qquad
    \subfloat{{\includegraphics[width=7.2cm]{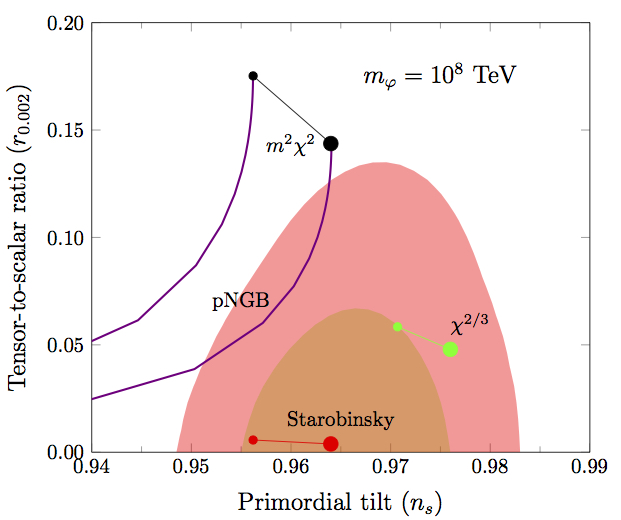} }}%
    \qquad
    \subfloat{{\includegraphics[width=7.2cm]{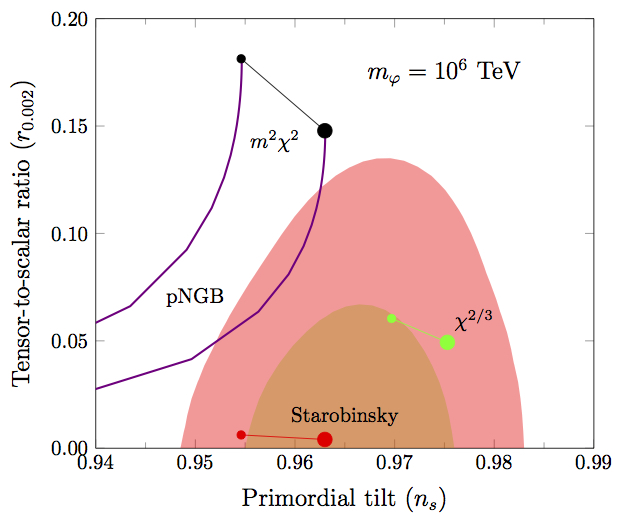} }}%
    \qquad 
     \subfloat{{\includegraphics[width=7.2cm]{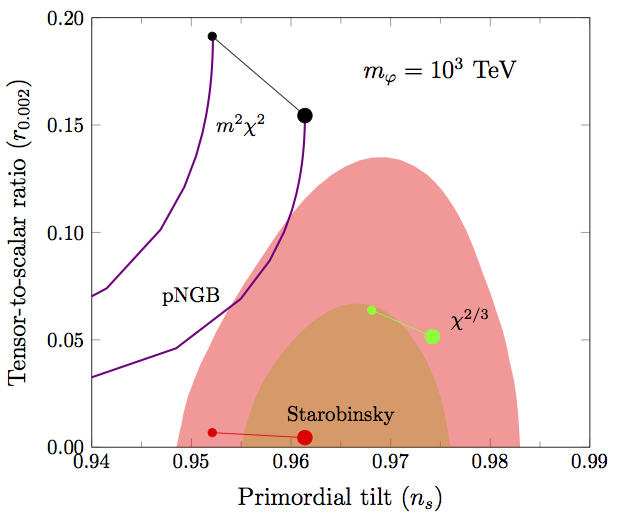} }}%
             \caption{Inflationary predictions for $m^{2} \chi^{2}$ (black), Natural/pNGB inflation (purple), Axion monodromy (green), Starobinsky model (red). For the cases of no misalignment $(\mvp > H_{\rm infl})$,  $m_{\vp} = 10^{3}, 10^{6},10^{8} \quad \rm{TeV.}$}%
    \label{assorted}%
\end{figure}

\noindent The results are shown 
in Figure \ref{assorted}, the plot for the standard cosmological timeline (which is equivalent to $\mvp >
H_{\rm infl}$) is also included for reference. The shaded regions correspond to  the $1\tiny{-}\sigma$ and $2\tiny{-}\sigma$ results for $n_s$ and $r$ from {\sc{planck}} 2015 analysis for TT modes and low P  \cite{planck15}. We find  that for the $m^2 \chi^2$ model even a very heavy modulus of mass $10^8 \spa \rm{TeV}$ implies predictions for $n_s$ and $r$ which are well outside the $2\tiny{-}\sigma$ region. The axion monodromy model moves inside the
$1\tiny{-}\sigma$ region for  $\mvp$ below $10^{5} \spa {\rm TeV}$. The Starobinsky model remains in the $1\tiny{-}\sigma$ region for almost the entire
mass range. 

  Finally, we would like to mention a general implication. For gravity mediated models 
moduli masses are tied to the scale of supersymmetry breaking. Thus, for gravity
mediated models our results correlate inflationary predictions with the scale of supersymmetry
breaking. The effect is significant even for models with a high scale of supersymmetry breaking.

\section{A Bound on Moduli Masses}

 The consistency condition \eqref{general3}  can be used to obtain a bound on moduli masses
given a model of inflation by taking input from observations on the value of $n_s$ \cite{dm}. The
approach can be considered complimentary to that of the previous section where we discussed inflationary predictions as a function of the mass of the late time decaying modulus. In this section,
we analyse the bound for our representative models and update some of the discussion in \cite{dm} in light
of the  {\sc{planck}} 2015 data release \cite{planck15}.

  We begin by briefly reviewing the derivation of the bound. Combining the expression for $N_{\rm mod}$
\pref{Modd} with the consistency condition \pref{general3} one finds
\bea
\! \frac{1}{3}\ln\! \! \left(\! \frac{\!\sqrt{16\pi} M_{\rm pl} Y^2\!}{m_{\varphi}}\!\right)\!\! + \!{1 \over 4}(\! 1\! -\! 3w_{\rm re1}\!) N_{\rm re1}\!  + \! {1 \over 4}( 1\! -\! 3w_{\rm re2}) N_{\rm re2}\! \approx \!55.43\! -\! N_k \! +\!\frac{1}{4}\ln r \!+ \!{ 1 \over 4 } \ln\! \left(\!\frac{{ \rho_{ k}}} {\rho_{\rm end}} \!\right)\!. \label{consistency_conditions_bound}
\eea
Various numerical and analytic studies of reheating suggest strongly that the  effective equation of state during reheating epochs is less that one third  
 i.e. $w_{\rm re1}, w_{\rm re2} < 1/3$ (see for e.g. \cite{Da, planck14} for a discussion). With this, the second and third term in the left hand side of 
\pref{consistency_conditions_bound} are  positive definite and the equation can immediately 
be converted to a bound for the mass of the modulus
 \bea
 \label{bound1}
 m_{\varphi}\gtrsim   \sqrt{16 \pi} M_{\rm pl} Y^2 ~e^{-3\left(55.43  - N_k   + { 1 \over 4 } \ln \left({ \rho_{\rm k} / \rho_{\rm end} } \right)+\frac{1}{4}\ln r\right)}.
 \eea
 The bound applies only if $\mvp$ is less than Hubble during inflation (as equation \pref{general3} was
 derived under this assumption). We have used $w_{\rm re1}, w_{\rm re2} < 1/3$, in arriving at the bound. In specific models,
 one can hope to compute the parameters $w_{\rm re1}, w_{\rm re2}, N_{\rm re1}$ and $N_{\rm re2}$.
 Equation \pref{consistency_conditions_bound} can then be regarded as relating the mass of the 
 modulus to the inflationary sector. Note that the longer the duration of reheating higher the value of
 $\mvp$.
 
  Given a model of inflation and observational input on the value of $n_s$, one can explicitly compute
the quantities in the exponent in the right hand side of \pref{bound1}. Typically, $N_k$ is related
to $n_s$  by a relation of the form
$$
   N_k = { \beta \over {1 - n_s} },
$$
where $\beta$ depends on the model of inflation. This makes the bound highly sensitive to the value 
of $n_s$.  The   {\sc{planck}} 2015 release \cite{planck15} gives the central value of $n_s$ to be 0.9680; there is a shift in the positive direction in comparison with the 2013 value of $n_s=0.9603$ \cite{planck14}.  This implies an increase in $N_k$ for inflationary models 
and thereby a more stringent bound. 

\begin{figure}[!h] 
   \vspace{-0.5cm}
    \begin{center}
    \includegraphics[scale=0.3]{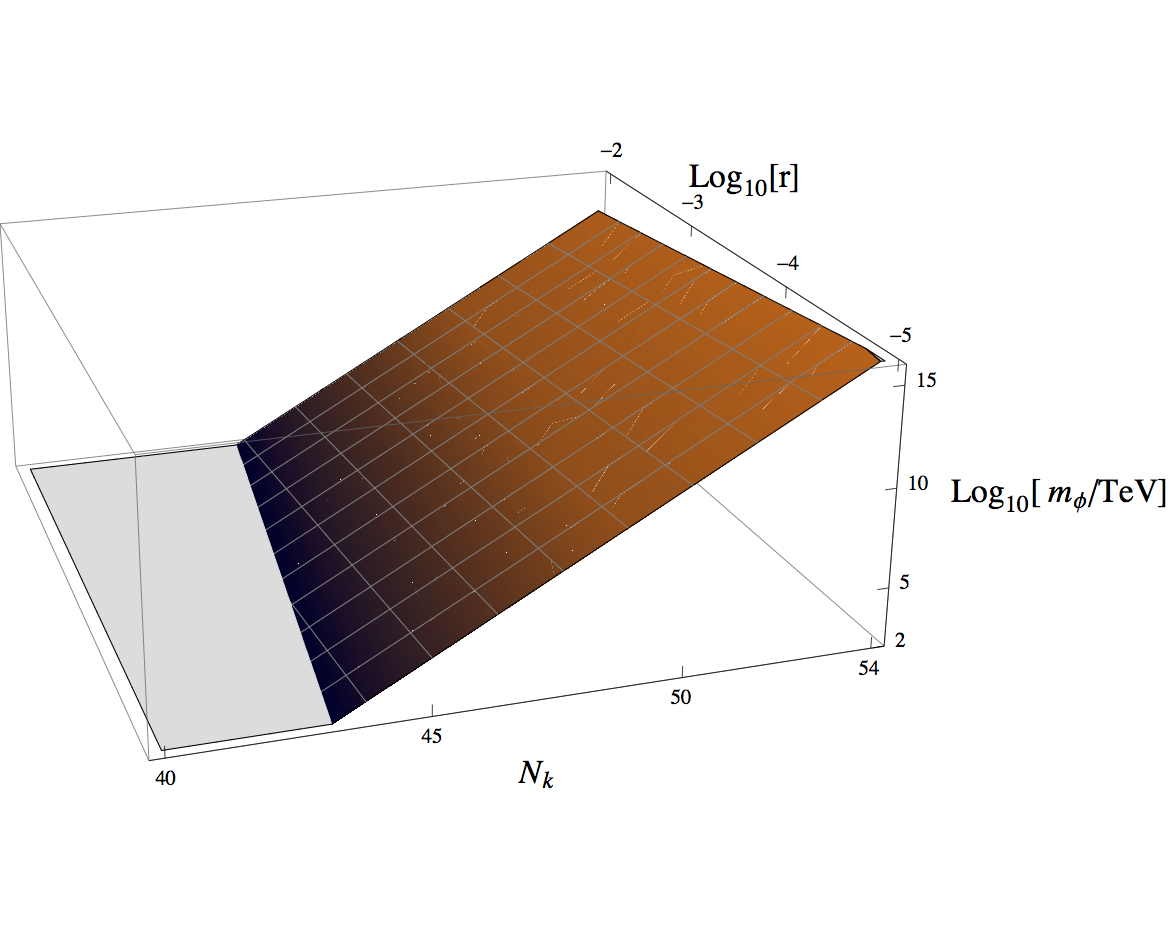}%
    \end{center}
    \vspace{-2cm}
             \caption{Bound on the modulus mass for small field models. The  allowed values of $\mvp$ are in the region
             above the shaded plane.  We have chosen $Y = 1/10$.}%
             \label{bound_plot}
\end{figure}

   Let us now discuss the bound in the context of our representative models. For polynomial potentials
$ V(\chi) \propto \chi^{\alpha}$, the bound simplifies to
\begin{equation}
\label{ACV}
m_{\varphi} \gtrsim \sqrt{16\pi} M_{\rm pl} Y^2 e^{-3(55.85 - \frac{(2 + \alpha)}{2(1 - n_s)})}.
\end{equation}
For the $m^2 \chi^2$ model, the {\sc planck} 2015 central value of $n_s$ gives the right hand side of \pref{ACV} to be well
above Hubble during inflation (as obtained in Figure 1); modular cosmology is incompatible with this value of $n_s$. 
The lower end of the $1\tiny{-}\sigma$ value gives $m_{\vp} > 10^{10} \spa \rm{TeV}$. On the other hand, for the axion monodromy model $(\alpha =2/3)$ \pref{ACV} yields a value below the
CMP bound \pref{bound}, thus is not of phenomenological interest as a bound. The fact that the bound is not strong for the axion monodromy model
is consistent with the results shown in figure 2 - the axion monodromy model is in the $1\tiny{-}\sigma$
region for $m_{\vp} = 10^3 \spa \rm{TeV}$.
Similarly, in the case of the Starobinsky model and pNGB inflation the value of the bound is in keeping with
the results shown in figure 2.

   For small field models, the second term in the exponent of the right hand side of  \pref{bound1}
(the term involving the ratio of the energy densities at the time of horizon exit and end of inflation)
makes a negligible contribution.  Note that the functional form of the bound is such that it becomes
stronger with decreasing $r$. In Figure \ref{bound_plot} we show the allowed range for $\mvp$
as a function of $N_k$ and $r$. The plot illustrates that the scale for the bound is essentially set by $N_k$. 
For $N_k \gtrsim 50$ the bound is very strong; $m_{\varphi} \gtrsim 10^{7} \spa {\rm TeV}$.
The bound is stronger than the CMP bound \pref{bound} as long as $N_k \gtrsim 44.5$. The plot in Figure
\ref{bound_plot} can be used to read off the implications of the bound for any small field model. It will
be interesting to explore the implications of this bound for inflationary model building in moduli
stabilised string compactifications.

\section{Discussion and Conclusions}

   In this paper, we have studied the sensitivity of $n_s$ and $r$ to the mass of the lightest
modulus in the context of modular cosmology. The results of section 3 clearly exhibit that
it is important to explicitly incorporate the effect of the epoch of modulus domination in 
obtaining the preferred range of $N_k$. The effect can significantly alter the inflationary
predictions for $n_s$ and $r$ of string/supergravity models; being relevant even for very heavy moduli $(m_{\varphi} 
\approx 10^{8} \spa \rm{TeV})$. Furthermore, future experiments \cite{euclid} are likely to
bring down the uncertainties in the measurement of $n_s$ by one order of magnitude; making our
analysis all the more relevant. Given that modular cosmology is generic in string/supergravity 
models \cite{cmp,ccmp,cmmp,douglas, bobby, fq} our results should have broad implications.

   Our approach has been phenomenological; we have treated the mass of the lightest modulus as
a free parameter and taken the initial displacement of the modulus (that results due to misalignment)
to have a generic value. The results strongly motivate the study of specific models where the modulus
mass takes a fixed value and it is possible to compute the value of the initial displacement
explicitly. Some  models worth exploring in this context
are fibre inflation  \cite{finer} and Kahler moduli inflation \cite{kahler}. 

   Another important direction in the study of specific models is  first principles analysis of the reheating
epoch. This can reduce the uncertainty in $N_k$, allowing for more precise predictions of $n_s$
and $r$. This question has received much attention recently  
\cite{Da,AB,CD}. The methods developed in \cite{CD} can be useful in analysing the
decay of moduli particles. 

   More generally, modular cosmology can also have implications for dark matter, structure
formation and the phenomenology of SUSY models \cite{Pheno}. It is natural to look for correlations
between our results for CMB observables and other phenomenological signatures. Gravity  mediated
models  are particularly interesting in this context, as the moduli masses are tied to the
scale of supersymmetry breaking.

\section*{Acknowledgements}

We would like to thank Luis Aparicio, Mar Bastero-Gil, Michele Cicoli, Joseph Conlon, John Ellis, Henriette Elvang, Lucien Heurtier, Gordy Kane, Sven Krippendorf, Fernando Quevedo, 
Raghu Rangarajan, Gary Shiu, Kuver Sinha, Mark Srednicki, Fuminobu Takahashi, Clemens Wieck, Scott Watson and Ivonne Zavala  for useful discussions. K Das would like to thank the Harish Chandra Research Institute
for hospitality. K Dutta 
would like to acknowledge support from a Ramanujan fellowship of the Department of Science and 
Technology, India and Max Planck Society-DST Visiting Fellowship grant. K Dutta would also like to thank Max Planck Institute for Physics, Munich, where a part of the work is completed. AM
would like to acknowledge support from a Ramanujan fellowship of the Department of Science and Technology, India. AM would like to thank the Hong Kong University of Science and Technology, KEK Theory Group, Michigan Centre for Theoretical Physics, University of California at Santa Barbara and
University of Swansea for hospitality.

\section*{Appendix}

\subsection*{A. Density Perturbations in Modular  Cosmology}

       Our focus has been on models in which  quantum fluctuations during the
inflationary epoch are responsible for the density perturbations. Here we elaborate on this further
in the context of modulus dominated cosmology. As discussed in section \ref{Modcos} the minimum of the potential of the late time decaying modulus depends on the inflaton expectation value; thus as the inflaton moves along
its trajectory the expectation value of the late time decaying modulus (and potentially other
moduli) necessarily changes. Thus, the trajectory in field space during inflation involves displacement
along the inflaton direction, late time decaying modulus (and potentially other moduli).  We will require the directions in field space orthogonal to the trajectory in field space during inflation to have mass of at least of the order of Hubble
(this as we will see in what follows will ensure that isocurvature perturbations are suppressed). Infact, curvature couplings naturally lead to such mass terms of the order of Hubble (see for e.g.\cite{DineR, DineRT}).

The perturbations generated are best understood
in the formalism developed in \cite{wand1} --    coordinates in field space are chosen such that one of the
coordinate directions is along the trajectory in field space (during the inflationary epoch) and the remaining are   orthogonal to the trajectory in field space. The key result of \cite{wand1} is that quantum fluctuations associated with the direction in field space parallel to the trajectory are adiabatic, while the
ones orthogonal generate isocurvature perturbations. Thus,  imposing the condition that
the directions in field space orthogonal to the trajectory have mass at least of the order of
Hubble ensures that isocurvature perturbations at the time of horizon exit are suppressed; the perturbations are to a very good approximation adiabatic at the time of horizon exit. We will denote the adiabatic perturbation at the time
of horizon exit by $\cal{R}_{*}$ and the isocurvature perturbations by ${\cal{S}}^{i}_{*}$. These 
have to be evolved into the radiation epoch (after the decay of the modulus) to determine the strength of
the temperature fluctuations they seed. The result of this evolution is given by a transfer matrix 
\cite{transfer}, which takes the general form (to keep the presentation simple
we include one isocurvature direction, it is easily generalised to the case of multiple isocurvature perturbation directions)
\[
\begin{bmatrix}
\mathcal{R}_{\rm rad} \\
\mathcal{S}_{\rm rad}
\end{bmatrix}
=
\begin{bmatrix}
1 & \mathcal{T}_{\rm RS}\\
0 & \mathcal{T}_{\rm SS}
\end{bmatrix}
\begin{bmatrix}
\mathcal{R}_{*} \\
\mathcal{S}_{*}
\end{bmatrix}
\]
where $\mathcal{R}_{\rm rad}$ and $\mathcal{S}_{\rm rad}$  are the isocurvature and adiabatic perturbations
after the modulus decay. An important feature of the transfer matrix is that the entries in the first column 
are completely model independent \cite{transfer} - they follow from the fact that a purely adiabatic perturbation is conserved
and does not lead to any isocurvature  perturbations. On the other hand, the transfer functions $\mathcal{T}_{\rm RS}$ and $\mathcal{T}_{\rm SS}$ are model dependent. But, the form of the transfer matrix implies that if $\mathcal{S}_{*} << \mathcal{R}_{*}$, then isocurvature perturbations remain suppressed and $\mathcal{R}_{\rm rad}$ is essentially determined
by $\mathcal{R}_{*}$. Thus, for models in which the  only light direction during the inflationary 
epoch is the trajectory in field space the density perturbations are adiabatic and determined by the curvature 
perturbation at the time of horizon exit.

  Other scenarios to generate density perturbations are the curvaton scenario \cite{CU} and modulated fluctuations
  \cite{modulated}.  We shall not explore these possibilities here, see \cite{ZB,ZB2} for their realisations
in  string models.

   \end{document}